\documentclass[journal,article,submit,pdftex,moreauthors]{Definitions/mdpi} 
\firstpage{1} 
\pubvolume{1}
\issuenum{1}
\articlenumber{0}
\pubyear{2022}
\copyrightyear{2022}
\datereceived{} 
\dateaccepted{} 
\datepublished{} 
\hreflink{https://doi.org/} 

\Title{Paramerization of deceleration parameter in $f(Q)$ gravity}

\TitleCitation{Paramerization of deceleration parameter in $f(Q)$ gravity}

\Author{Gaurav N. Gadbail $^{1}$\orcidA{}, Sanjay Mandal $^{2}$\orcidB{} and P.K. Sahoo $^{3,}$*\orcidC{}}

\AuthorNames{Gaurav N. Gadbail, Sanjay Mandal and P.K. Sahoo}

\AuthorCitation{Gadbail, G.N.; Mandal, S.; Sahoo, P.K.}

\address{%
$^{1}$\quad Department of Mathematics, Birla Institute of Technology and
Science-Pilani, Hyderabad Campus, Hyderabad-500078, India.; gauravgadbail6@gmail.com\\
$^{2}$\quad Department of Mathematics, Birla Institute of Technology and
Science-Pilani, Hyderabad Campus, Hyderabad-500078, India.; sanjaymandal960@gmail.com\\
$^{3}$\quad Department of Mathematics, Birla Institute of Technology and
Science-Pilani, Hyderabad Campus, Hyderabad-500078, India.; pksahoo@hyderabad.bits-pilani.ac.in}

\corres{Correspondence: pksahoo@hyderabad.bits-pilani.ac.in}

\abstract{In this article, we investigate the modified symmetric teleparallel gravity or $f(Q)$ gravity, where $Q$ is the non-metricity, to study the evolutionary history of the universe by considering the functional form of $f(Q)=\alpha Q^n$, where $\alpha$ and $n$ are constants. Here, we consider the parametrization form of the deceleration parameter as $q=q_0+\frac{q_1\,z}{(1+z)^2}$ which provides the desired property for sign flip from a decelerating to an accelerating phase.
We get the solution of the Hubble parameter by examining the mentioned parametric form of $q$, and then we impose the solution in Friedmann equations. Employing the Bayesian analysis for the Observational Hubble data (OHD), we estimated the constraints on the associated free parameters $(H_0,q_0,q_1)$ to determine if this model may challenge the $\Lambda$CDM limitations. Furthermore, the constrained current value of the deceleration parameter $q_0=-0.832^{+0.091}_{-0.091}$ shows that the present universe is accelerating. We also investigate the evolutionary trajectory of energy density, pressure, and EoS parameters to conclude the accelerating behavior of the universe. Finally, we try to demonstrate that the considered parametric form of the deceleration parameter is compatible with $f(Q)$ gravity. }


\begin{document}


\section{Introduction}
Recently, several cosmological observations \cite{Perlmutter/1999,Riess/1998,Riess/2004,Spergel/2007,Koivisto/2006,Daniel/2008} have supported the late-time cosmic acceleration expansion of the universe. However, based on the same cosmological observation, it is estimated that dark energy (DE) and dark matter (DM) cover up $95-96\%$ of the universe's composition, comprising mysterious dark components, so-called dark matter and dark energy, whereas baryonic matter covers up $4-5\%$ of the content of the universe. Presently, General relativity is believed to be the most successful theory of gravitation and its few gravitational tests have been discussed in \cite{corda}. However, it cannot provide a satisfactory explanation for the dark energy and dark matter problem; hence, it may not be regarded as the ultimate gravitational force theory for dealing with current cosmological problems. Several alternative approaches have been proposed in the literature over the last several decades to overcome the current cosmological problems. Nowadays, the modified theory of gravity is the most admirable candidate to solve the current difficulties (DE and DM problem) of the universe. One of the most prominent schemes to address the dark content issue of the universe is the modification of GR called the $f(R)$ theory of gravity, where $R$ is the Ricci scalar \cite{Buchdahl/1970}. Some other modified theories are also developed to solve this issue such as $f(\mathcal{T})$ theory, where $\mathcal{T}$ is the torsion \cite{Ferraro/2007,Myrzakulov/2011}, $f(R,T)$ theory \cite{Harko/2011,Sahoo/2018}, $f(R,L_m)$ theory, where $L_m$ is the matter Lagrangian density \cite{Harko/2010,Jaybhaye/2022}, $f(R,G)$ theory, where $G$ is the Gauss-Bonnet invariant \cite{Elizalde/2010,Bamba/2010} and many more.\\

Jimenez et al. \cite{Jimenez/2018} recently proposed a novel proposal by considering a modification of Symmetric Teleparallel Equivalent to GR called $f(Q)$ gravity, where $Q$ is a non-metricity scalar. The non-metricity Q of the metric geometrically characterizes the variation of the length of a vector in parallel transport, and it represents the primary geometric variable explaining the features of gravitational interaction. Recently,  several studies have been done on $f(Q)$ gravity. Mandal et al. studied cosmography \cite{Mandal/2020a} and energy condition \cite{Mandal/2020b} in nonmetric $f(Q)$ gravity. For the purpose of examining an accelerated expansion of the universe with recent observations, Lazkoz et al. \cite{Lazkoz/2019} examined several $f(Q)$ gravity models. Furthermore, Solanki et al. \cite{Solanki/2021} studied the effect of bulk viscosity in the accelerating expansion of the universe in $f(Q)$ gravity. Esposito et al. \cite{Esposito/2022} examined exact isotropic and anisotropic cosmological solutions using reconstruction techniques. Moreover, $f(Q)$ gravity easily overcomes the limits set by Big Bang Nucleosynthesis (BBN) \cite{Anagnostopoulos/2022}. Many other works have been completed within the context of $f(Q)$ gravity theory \citep{Harko/2018,Frusciante/2021,Khyllep/2021,Ayuso/2021,GG/2022,Capo/2022}.\\
Although various theoretical approaches exist to explain the phenomenon of cosmic acceleration, none is definitively known as the appropriate one. The current model of late-time cosmic acceleration is known as reconstruction. This is the inverse method of locating a suitable cosmological model. There are two kinds of reconstruction: parametric reconstruction and non-parametric reconstruction. The parametric reconstruction relies on estimating model parameters from various observational data. It is also known as the model-dependent approach. The main idea is to assume a specific evolution scenario and then determine the nature of the matter sector or exotic component that is causing the alleged acceleration. Several authors have used this method to find a suitable solution \cite{Mukherjee/2016,Arora/2021a,Gadbail/2022}.

In this work, we consider the parametrization form of the deceleration parameter in terms of redshift $z$ as $q(z)=q_0+\frac{q_1 z}{(z+1)^2}$ which provides the desired property for sign flip from a decelerating to an accelerating phase, and investigate the FLRW universe in the framework of nonmetric $f(Q)$ gravity by using the functional form of $f(Q)$ as $f(Q)=\alpha Q^n$, where $\alpha$ and $n$ are arbitrary constants. The present paper is arranged as follows: In Section \ref{section 2}, we started with basic $f(Q)$ gravity formalism and developed the field equation for the FLRW line element. In section \ref{section 3}, we adopting the parametric form of deceleration parameter and then find the Hubble solution. In section \ref{section 4},  we estimated the constraints on the associated free parameters $(H_0,q_0,q_1)$ by employing the Bayesian analysis for the Observational Hubble data (OHD). Then we check the evolutionary trajectory of energy density, pressure, and EoS parameters to conclude the accelerating behavior of the universe in section \ref{section 5}. Lastly, we conclude our result in section \ref{section 6}.

\section{$f(Q)$ gravity Formalism}
\label{section 2}
The most generic action of nonmetric $f(Q)$ gravity is given by \cite{Jimenez/2018}
\begin{equation}
\label{1}
S=\int \left[\frac{1}{2}f(Q)+\mathcal{L}_m\right]\sqrt{-g}d^4x,
\end{equation}
where $f$ is an arbitrary function of non-metricity scalar $Q$,
 $\mathcal{L}_m$ is the matter Lagrangian density, and $g$ is a determinant of the metric tensor $g_{\alpha\beta}$. \\
 The definition of non-metricity tensor in $f(Q)$ gravity is 
\begin{equation}
 Q_{\sigma\alpha\beta}=\nabla_{\sigma}\,g_{\alpha\beta}
\end{equation}  
 
 and the corresponding  traces are 
 
\begin{equation}
\label{2}
Q_{\sigma}=Q_{\sigma\,\,\,\,\alpha}^{\,\,\,\,\alpha}\, ,\,\,\,\,\,\,\,\,\tilde{Q}_{\sigma}=Q^{\alpha}_{\,\,\,\,\sigma\alpha}\,.
\end{equation}
Moreover, the superpotential tensor $P_{\,\,\mu\nu}^{\lambda}$ is given by
\begin{equation}
\label{3}
4P_{\,\,\alpha\beta}^{\sigma}=-Q^{\sigma}_{\,\,\,\,\alpha\beta}+2Q^{\,\,\,\,\,\,\sigma}_{(\alpha\,\,\,\,\beta)}-Q^{\sigma}g_{\alpha\beta}-\tilde{Q}^{\sigma}g_{\alpha\beta}-\delta^{\sigma}_{(\alpha}\, Q\,_{\beta)},
\end{equation} 
Hence, the nonmetricity scalar can be obtained as
\begin{equation}
\label{4}
Q=-Q_{\sigma\alpha\beta}P^{\sigma\alpha\beta}.
\end{equation}

The gravitational field equation derived by varying the action \eqref{1} with regard to the metric tensor is presented below:

\begin{equation}
\label{6}
\frac{2}{\sqrt{-g}}\nabla_{\sigma}\left(f_{Q}\sqrt{-g}\,P^{\sigma}_{\,\,\alpha\beta}\right)+\frac{1}{2}f\,g_{\alpha\beta}+
f_{Q}\left(P_{\alpha\sigma\lambda}Q_{\beta}^{\,\,\,\sigma\lambda}-2Q_{\sigma\lambda\alpha}P^{\sigma\lambda}_{\,\,\,\,\,\,\beta}\right)=- T_{\alpha\beta},
\end{equation}
where $T_{\alpha\beta}\equiv-\frac{2}{\sqrt{-g}}\frac{\delta(\sqrt{-g})\mathcal{L}_m} {\delta g^{\alpha\beta}}$ and  $f_Q=\frac{d f}{d Q}$.\\ Similarly, by varying the action \eqref{1} with regard to the connection, the following result can be obtained:
\begin{equation}
\label{7}
\nabla_{\alpha}\nabla_{\beta} \left(f_{Q}\sqrt{-g}\,P_{\,\,\,\,\,\,\,\sigma}^{\alpha\beta}\right)=0.
\end{equation}

We shall consider a spatially flat FLRW universe throughout the investigation, whose metric is given by
\begin{equation}
\label{8}
ds^2=-dt^2+a^2(t)(dx^2+dy^2+dz^2).
\end{equation}
Here, $a(t)$ is a cosmic scale factor. The non-metricity scalar $Q=6H^2$ obtained for the above FLRW metric, where $H=\frac{\dot{a}}{a}$ is the Hubble parameter. In this case, the energy-momentum tensor of a perfect fluid $T_{\alpha\beta}=(p+\rho)u_{\alpha}u_{\beta}+pg_{\alpha\beta}$, where $p$ and $\rho$ are pressure and energy density, respectively.\\ For the metric \eqref{8}, the corresponding Friedmann equations are \cite{Jimenez/2018}: 
\begin{equation}
\label{9}
3 H^2=\frac{1}{2f_Q}\left(-\rho+\frac{1}{2}f\right),
\end{equation} 
\begin{equation}
\label{10}
\dot{H}+3H^2+\frac{\dot{f}_Q}{f_Q}H=\frac{1}{2f_Q}\left(p+\frac{1}{2}f\right).
\end{equation}

Using the preceding Friedmann equations in the context of $f(Q)$ gravity, one may now study possible cosmological applications.
\section{Parametrization of the deceleration parameter}
\label{section 3}

The parametrization of the deceleration parameter $q$ plays a significant role in determining the nature of the universe's expanding rate. In this regard, some research employed various parametric forms of deceleration parameters, while some others investigated non-parametric forms. These methods have been widely discussed in the literature to characterise concerns with cosmological investigations, such as the initial singularity problem, the problem of all-time decelerating expansion, the horizon problem, Hubble tension, and so on \cite{Banerjee/2005,Cunha/2008,Escamilla/2022}. Motivated by this fact, in this article, we consider simplest parametric form of the deceleration parameter $q$ in terms of redshift $z$ as \cite{Wang/2010}
\begin{equation}
\label{11}
q(z)=q_0+\frac{q_1\, z}{(z+1)^2}
\end{equation}
where $q_0=q(z=0)$ indicates the present value of deceleration parameter, and $q_1$ depicts the variation in the deceleration parameter as a function of $z$. Clearly, one of the most well-liked parametrizations of the dark energy equation of state served as inspiration for this parametric form for $q(z)$ \cite{Gong/2006}, and it seems to be versatile enough to fit the $q(z)$ behavior of a broad class of accelerating models.\\
 The derivative of the Hubble parameter with respect to time $t$ is $\dot{H}=-(1+q)H^2$. Then there exists a relation between the Hubble parameter and the deceleration parameter in virtue of an integration:
\begin{equation}
\label{12}
H(z) = H_0\,\, \text{exp}\left[\int_0^z (1+q(x))d ln(1+x)\right]
\end{equation}
where $x$ is a changing variable.
By using Eq. \eqref{11} in Eq. \eqref{12}, we obtained the Hubble parameter in terms of redshift $z$ as

\begin{equation}
\label{13}
H(z)=H_0 (z+1)^{q_0+1} e^{\frac{q_1 z^2}{2 (z+1)^2}}
\end{equation}
where $H_0$ is the Hubble constant.
Furthermore, utilising the relationship between redshift and the universe's scale factor $a(t)=\frac{1}{1+z}$, we may describe the relationship between cosmic time and redshift as
\begin{equation}
\label{14}
\frac{d}{dt}=\frac{dz}{dt}\frac{d}{dz}=-(1+z) H(z)\frac{d}{dz}
\end{equation}
Using Eq. \eqref{13} and \eqref{14} in Fridemann equations, we obtained the energy density $\rho$, pressure $p$ and equation of state paramater $\omega$ in terms of redshift $z$ as
\begin{equation}
\rho=\alpha  \left(-2^{n-1}\right) 3^n (2 n-1) \left(H_0^2\,(z+1)^{2 q_0+2} \,\,e^{\frac{q_1 z^2}{(z+1)^2}}\right)^n,
\end{equation}

\begin{multline}
p=\alpha  6^{n-1} \left(H_0^2 (z+1)^{2 q_0+2} e^{\frac{q_1 z^2}{(z+1)^2}}\right)^n
 \left(-\frac{2 n \left(q_0 (z+1)^2+z (q_1+z+2)+1\right)}{(z+1)^2}\right.\\
 \left.-\frac{4 (n-1) n (z+1)^{-q_0-3}\,\, e^{-\frac{q_1 z^2}{2 (z+1)^2}} \left(q_0 (z+1)^2+z (q_1+z+2)+1\right)}{H_0}+6 n-3\right),
\end{multline}

\begin{equation}
w=-\frac{-\frac{4\,n\,(n-1) (z+1)^{(-q_0-3)}\,\, e^{-\frac{q_1 z^2}{2 (z+1)^2}} \left(q_0 (z+1)^2+z (q_1+z+2)+1\right)}{H_0}-\frac{2 n \left(q_0 (z+1)^2+z (q_1+z+2)+1\right)}{(z+1)^2}+6 n-3}{3 (2 n-1)},
\end{equation}
respectively.
The behaviour and essential cosmological properties of the model described in equation \eqref{11} are wholly dependent on the model parameters $(q_0, q_1)$. In the next section, we constraint the model parameter $(H_0,q_0, q_1)$ by using the recent observational datasets to investigate the behavior of cosmological parameters.
\section{Observational Constraints and Cosmological Applications}
\label{section 4}

Now, one can deal with the various observational datasets to constraint the parameters $H_0,$\,\, $q_0$,\,\, $q_1$. In order to do study the observational data, we use the standard Bayesian technique and to obtain the posterior distributions of the parameters, we we employ a Markov Chain Monte Carlo (MCMC) method. Also, we use the \textit{emcee} package to do MCMC analysis. Here, in this study, we used the Hubble measurements ( i.e., Hubble data) to complete the stimulation. The following \textit{likelyhood} function is used to find the best fits of the parameters;
\begin{equation}
\label{21}
\mathcal{L} \propto exp(-\chi ^2/2),
\end{equation}
where $\chi^2$ is the \textit{pseudo chi-squared function} \cite{baye}. The $\chi^2$ functions for various dataset are discussed below.

\subsection{Cosmic Chronometer(CC) Sample}
\begin{center}
\begin{figure}[H]
\includegraphics[scale=0.7]{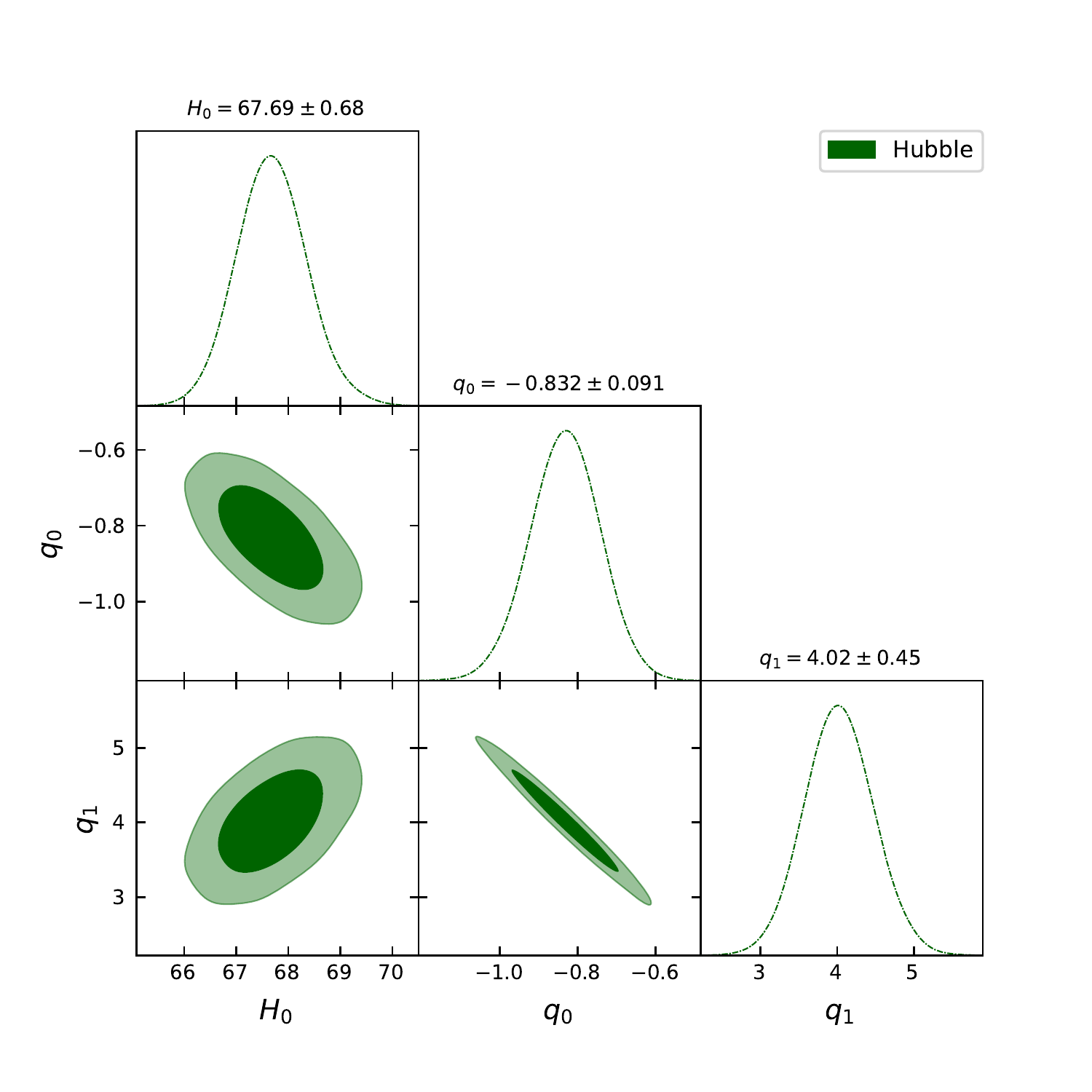}
\caption{The marginalized constraints on the coefficients in the expression of Hubble parameter $H(z)$ in Eqn. \eqref{13} are shown by using the Hubble sample.}
\label{f1}
\end{figure}
\end{center}

\begin{center}
\begin{figure}[H]

\includegraphics[scale=0.7]{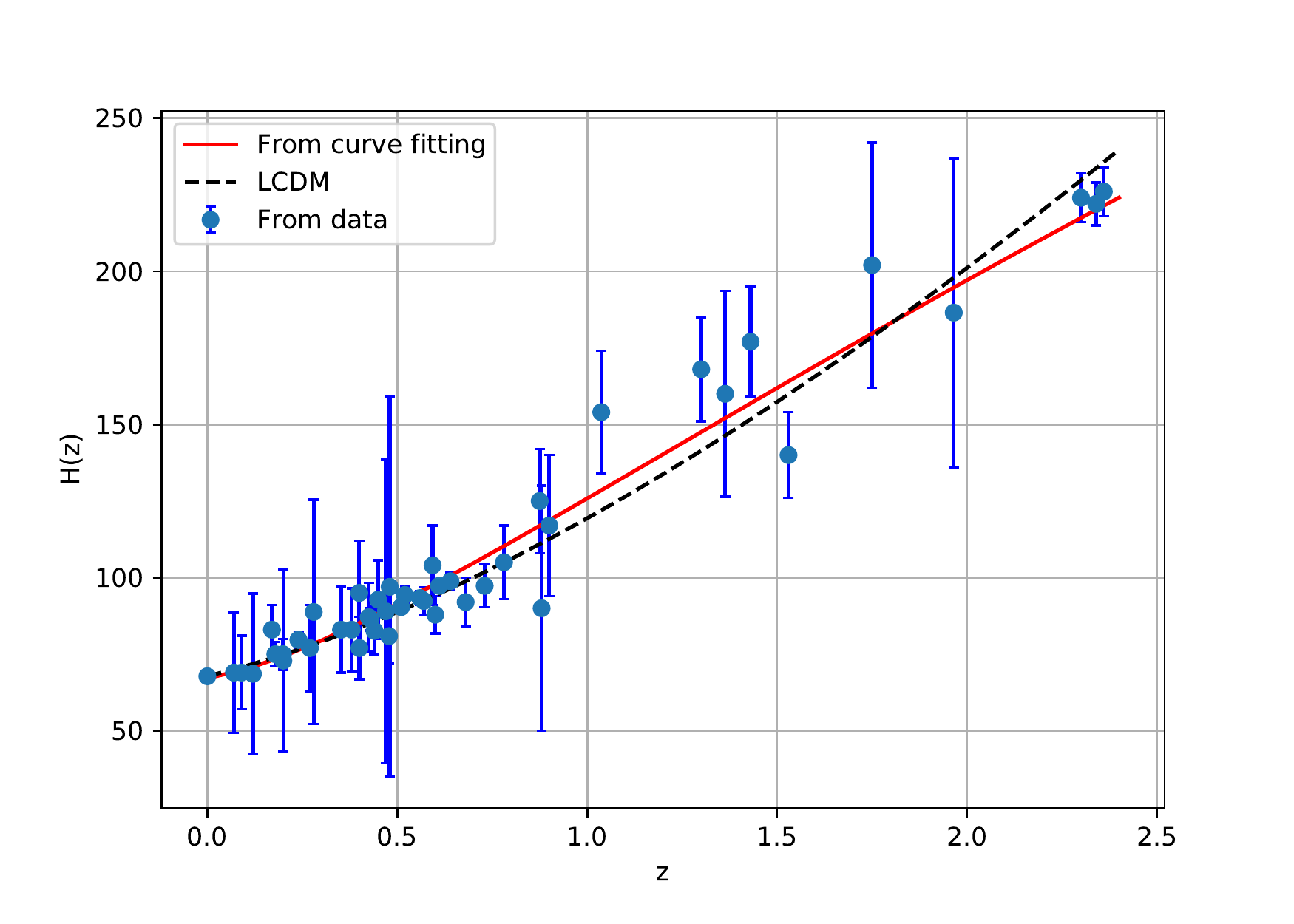}
\caption{The evolution of Hubble parameter $H(z)$ with respect to redshift $z$ is shown here. The red line represents our model and dashed-line indicates the $\Lambda$CMD model with $\Omega_{m0}=0.3$ and $\Omega_{\Lambda 0}=0.7$. The dots are shown the Hubble dataset with error bar.}
\label{f2}

\end{figure}
\end{center}

Recently, a list of Hubble measurements in the redshift range $0.07\leq z\leq 1.965$ were compiled by Singirikonda and Desai \cite{Desai/2020}. This H(z) dataset was measured from the differential ages $\Delta t$ of galaxies \cite{h1,h2,h3,h4}. The complete list of datasets is presented in \cite{Desai/2020}. To estimate the model parameters, we use the Chi-square function which is given by
\begin{equation}
\chi^2_{CC}(p_s)=\sum_{i=1}^{31}\frac{[H_{th}(p_s,z_i)-H_{obs}]^2}{\sigma^2_{H(z_i)}},
\end{equation}
where $H_{th}(p_s,z_i)$, $H_{obs}(z_i)$ represents the Hubble parameter with the model parameters, observed Hubble parameter values,respectively. $\sigma^2_{H(z_i)}$ is the standard deviation obtained from observations.
In Fig. \ref{f2}, the profile of our model against Hubble data shown. The marginalized constraining results are displayed in Fig. \ref{f1}.

\section{Cosmological parameters}
\label{section 5}
One of the cosmological parameters that is significant in explaining the state of expansion of our universe is the deceleration parameter $q$. When the value of the deceleration parameter is strictly less than zero shows the accelerating behavior of the universe, and when it is non-negative, the universe decelerates.  Furthermore, the observational data employed in this article revealed that our current universe is in an accelerating phase, with the present value of the deceleration parameter becoming $q_0=-0.832^{+0.091}_{-0.091}$. This type of result seen in the existing literature \cite{Mamon/2016,Hanafy/2019}.\\

\begin{figure}[H]

\includegraphics[scale=0.7]{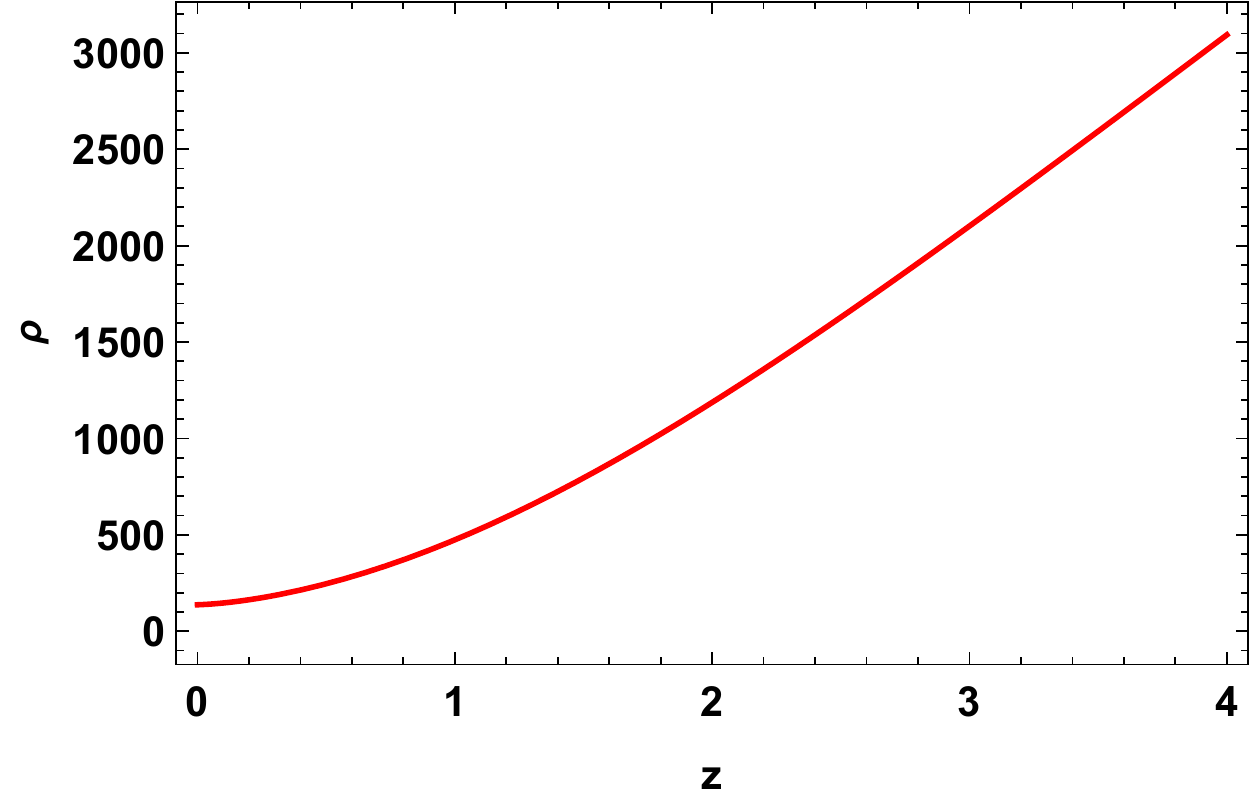}
\caption{Evolution trajectory of $\rho$ versus $z$ with constraint  values from Hubble datasets and $\alpha=-0.01$, $n=1.2$.}
\label{rho}

\end{figure}

\begin{figure}[H]

\includegraphics[scale=0.7]{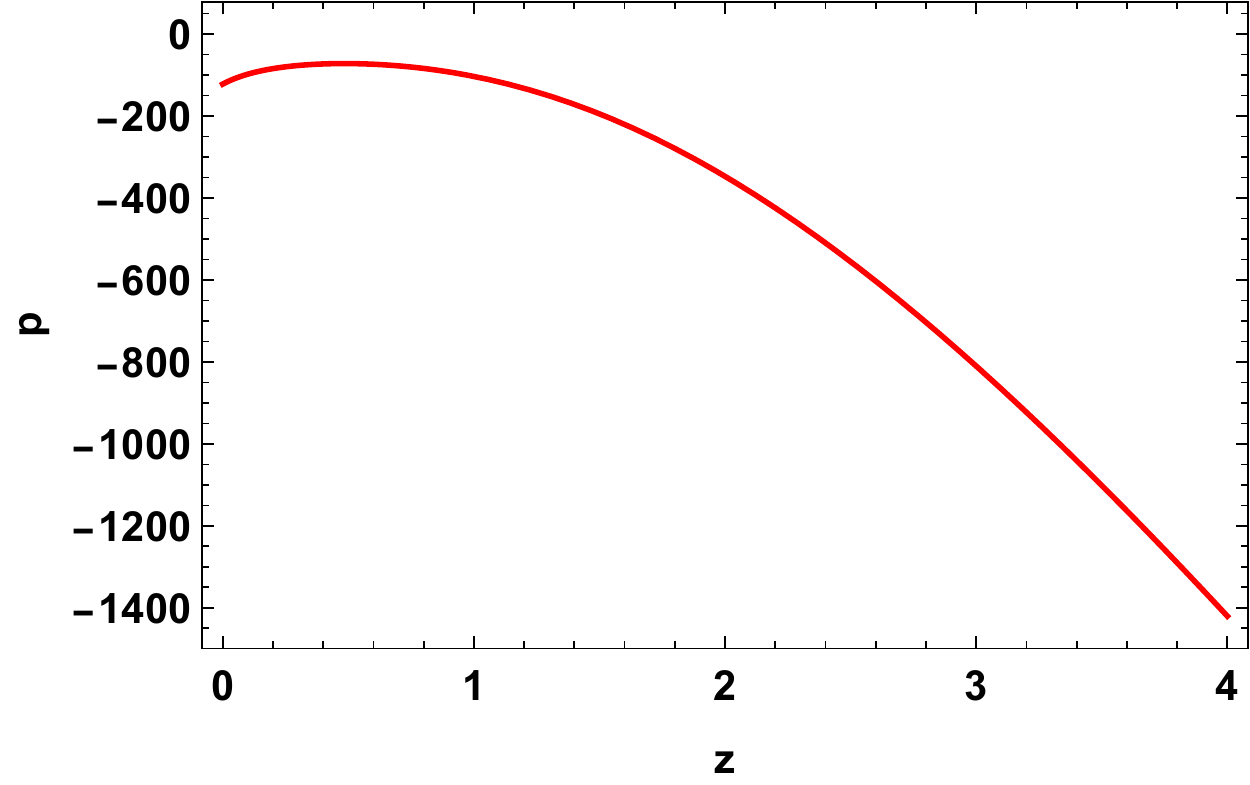}
\caption{Evolution trajectory of $p$ versus $z$ with constraint  values from Hubble datasets and $\alpha=-0.01$, $n=1.2$.}
\label{p}

\end{figure}

\begin{figure}[H]

\includegraphics[scale=0.7]{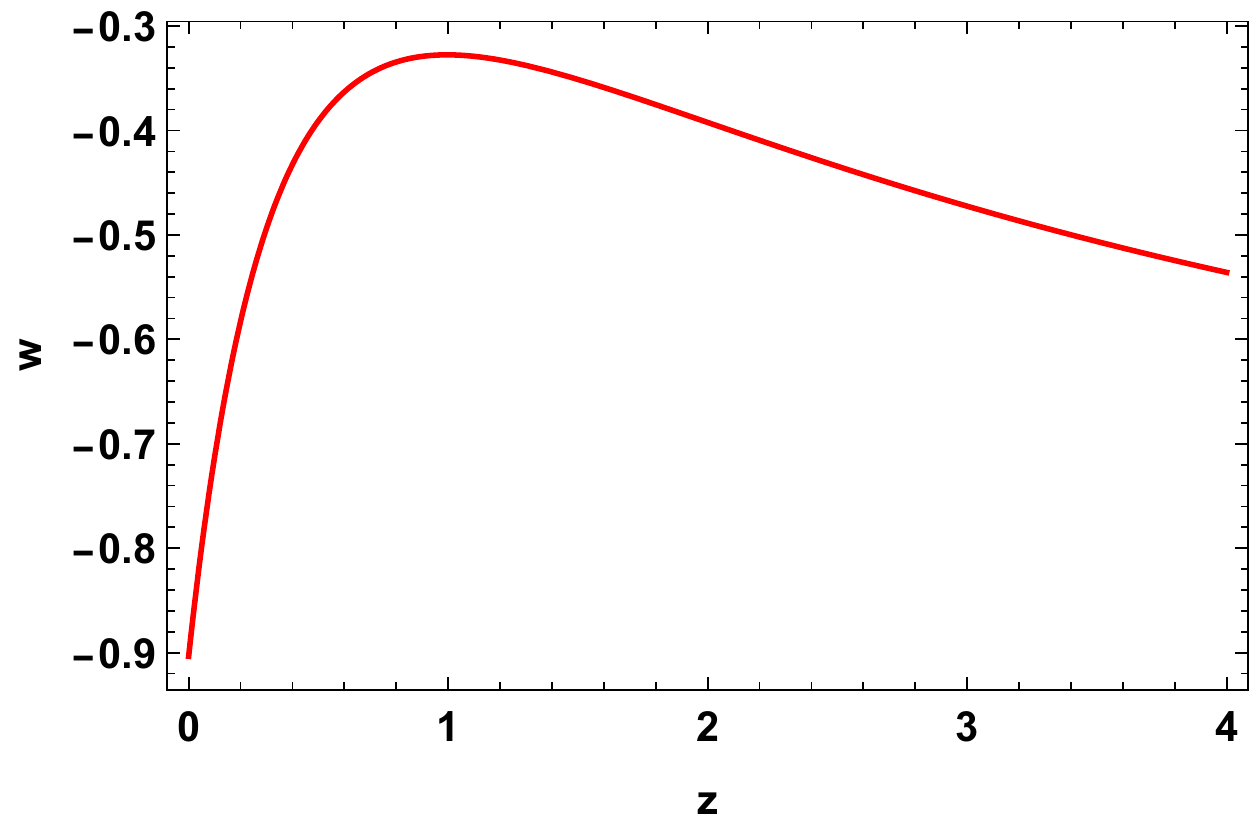}
\caption{Evolution trajectory of $w$ versus $z$ with constraint  values from Hubble datasets and $\alpha=-0.01$, $n=1.2$.}
\label{w}

\end{figure}

Figure \eqref{rho} indicates that the energy density of the universe increases with redshift and still seems as the universe expands, but Figure \eqref{p} demonstrates that the pressure decreases with redshift and has large negative values throughout cosmic evolution. The present cosmic acceleration induces this isotropic pressure behavior.\\
The EoS parameter $w$ is also helpful in categorizing the decelerating and accelerating behavior of the universe, and it is defined as $w=\frac{p}{\rho}$. The EoS categorize three possible state for accelerating universe is the quintessence $(-1<w<-\frac{1}{3})$ era, phantom $(w<-1)$ era and the cosmological constant $(w=-1)$.
Figure \eqref{w} shows the evolutionary trajectory of the EoS parameter, and it can be seen that the whole trajectory lies in the quintessence era. From Figure \eqref{w} it is clear that $w<0$ and the current value of EoS parameter is $w_0=-0.9^{+0.08}_{-0.12}$. Our result aligned with some of the studies \cite{Gong/2007,Gadbail/2022}, which indicates an accelerating phase. 


.

\section{Conclusion}
\label{section 6}
The current scenario of the accelerated expansion of the universe has grown increasingly fascinating over time. Numerous dynamical DE models and modified gravity theories have been employed in various ways to find a suitable description of the accelerating universe. In this work, we explored the accelerated expansion of the universe by adopting the parametric form of deceleration parameter in the framework of $f(Q)$ gravity, where $Q$ is the non-metricity scalar depicted the gravitational interaction.\\
We have examine the functional form of $f(Q)$ as $f(Q)=\alpha Q^n$, where $\alpha$ and $n$ are arbitrary constants, and the parametrization form of deceleration parameter as $q=q_0+\frac{q_1\,z}{(1+z)^2}$, where $(q_0,q_1)$ are model parameters. By utilizing the above parametric form, we find out the solution of Hubble parameter as $H(z)=H_0 (z+1)^{q_0+1} e^{\frac{q_1 z^2}{2 (z+1)^2}}$. Furthermore, we used the Hubble datasets containing $31$ data points to determine the best fit values for the model parameters $(H_0,q_0,q_1)$ as $H_0=67.69\pm 0.68$, $q_0=-0.832\pm 0.091$ and $q_1=4.02\pm 0.45$. Here, the $q_0$ shows the current value of the deceleration parameter, which depicts that the present expansion of the universe is accelerating. We analyzed the evolution of the various cosmological parameters corresponding to these best-fit values of model parameters. The EoS parameter exhibits negative behavior and lies in the quintessence era, which depicts that the present universe is an accelerating phase. Figure \eqref{rho} indicates that the energy density of the universe increases with redshift and still seems as the universe expands, but Figure \eqref{p} demonstrates that the pressure decreases with redshift and has large negative values throughout cosmic evolution.
Lastly, we conclude that considered parametric form of deceleration parameter in the framework of $f(Q)$ gravity theory plays an important role in driving the universe's accelerated expansion. 


\vspace{6pt} 



\authorcontributions{: Conceptualization, G.N.G.; Data curation, S.M.; Formal analysis, P.K.S.; Investigation, G.N.G.; Methodology, G.N.G.; Project administration, P.K.S.; Software, G.N.G. and S.M.; Supervision, P.K.S.; Validation, P.K.S.; Visualization, S.M.; Writing-original draft, G.N.G.; Writing-review and editing, S.M. and P.K.S. All authors have read and agreed to the published version of the manuscript.}

\funding{This research received no external funding}

\institutionalreview{Not applicable.}

\informedconsent{Not applicable.}

\dataavailability{There are no new data associated with this article.} 

\acknowledgments{GNG acknowledges University Grants Commission (UGC), New Delhi, India for awarding Junior Research Fellowship (UGC-Ref. No.: 201610122060). S.M. acknowledges Department of Science \& Technology (DST), Govt. of India, New Delhi, for awarding INSPIRE Fellowship (File No. DST/INSPIRE Fellowship/2018/IF180676). PKS acknowledges CSIR, New  Delhi, India for financial support to carry out the Research project [No.03(1454)/19/EMR-II, Dt. 02/08/2019] and IUCAA, Pune, India for providing support through the visiting Associateship program.
}

\conflictsofinterest{The authors declare no conflict of interest.} 


\begin{adjustwidth}{-\extralength}{0cm}

\reftitle{References}

\end{adjustwidth}
\end{document}